\documentclass[amstex, preprint, eqsecnum, prd, aps]{revtex4}

\begin{document}

\title{Anomalous diffusion in quantum Brownian motion with colored noise.}
\author{G. W. Ford}
\affiliation{Department of Physics, University of Michigan,. Ann Arbor,
MI 48109-1120}
\author{R. F. O'Connell{\footnote{E-mail: oconnell@phys.lsu.edu}}}
\affiliation{Department of Physics and Astronomy, Louisiana State
University, Baton Rouge, LA 70803-4001}
\date{\today}

\begin{abstract}
Anomalous diffusion is discussed in the context of quantum Brownian motion
with colored noise. It is shown that earlier results follow simply and directly from
the fluctuation-dissipation theorem. The limits on the long-time dependence
of anomalous diffusion are shown to be a consequence of the second law of
thermodynamics. The special case of an electron interacting with the
radiation field is discussed in detail.  We apply our results to wave-packet spreading.
\end{abstract}

\maketitle

\section{Introduction}

There has been a recurrence of interest in the phenomenon of anomalous
diffusion. We refer to a sampling of the literature \cite
{klafter,andrade,lutz,tsallis,scullya,scully,benkert,schaufler}. There
have also appeared in the literature a variety of explanations for such
behavior. In particular, Tsallis and others \cite{tsallis} have proposed
that conventional Gibbs-Boltzmann statistics be generalized by extending the
definition of entropy. On the other hand, it has been shown that an
explanation using conventional statistics is possible if one uses a
Fokker-Planck equation with either fractional derivatives \cite{andrade} or
time-dependent diffusion coefficients \cite{benkert,schaufler}. In addition,
a study using path integral methods \cite{grabert87} showed that anomalous
diffusion appears in quantum Brownian motion with colored noise. 

Our purpose here is first of all to show that these same results follow directly from
the well known exact formula for the correlation function that can be obtained using the
fluctuation-dissipation theorem. In this section, we present general results which are
applicable to an arbitrary heat bath.  Then, in Secs. II to IV, we apply these results
to the Ohmic heat bath, colored noise and a radiation heat bath (QED), respectively. 
In particularly, we note that the second law of thermodynamics places limits on the
asymptotic behaviour.  The physically realizable example of a charged particle coupled
to the radiation field is given particular attention.  In Sec. V, we apply our
results to the spreading of a wave packet.  Of special interest is the fact that the
results in the QED case involve the bare mass.  Finally, in Sec. VI, we present our
conclusions.

Diffusion of a Brownian particle is characterized by the long time behavior
of the mean square displacement, which for motion during time $t$ is defined
to be 
\begin{equation}
s(t)=\left\langle [x(t)-x(0)]^{2}\right\rangle .  \label{1.1}
\end{equation}
Normal diffusion corresponds to a linear time-dependence for long times,
with the diffusion constant given by $D=\frac{1}{2}\lim_{t\rightarrow \infty
}\dot{s}(t)$. Anomalous diffusion is characterized by a deviation from this
linear time dependence of the form 
\begin{equation}
s(t)\sim t^{1+\gamma },\qquad -1<\gamma <1,  \label{1.2}
\end{equation}
with $\gamma =0$ being the case of normal diffusion. Our discussion will be
based on the familiar formula for the correlation function, obtained using
the fluctuation-dissipation theorem \cite{callen,ford88a} 
\begin{equation}
\frac{1}{2}\left\langle x(t)x(0)+x(0)x(t)\right\rangle =\frac{\hbar }{\pi }
\int_{0}^{\infty }d\omega \mathrm{Im}\{\alpha (\omega +i0^{+})\}\coth \frac{
\hbar \omega }{2kT}\cos \omega t.  \label{1.3}
\end{equation}
Here $\alpha $ is the generalized susceptibility, which for a Brownian
particle of mass $m$ is of the form 
\begin{equation}
\alpha (z)=\frac{1}{-mz^{2}-iz\tilde{\mu}(z)},  \label{1.4}
\end{equation}
where $\tilde{\mu}(z)$ is the Fourier transform of the memory function. For
our purposes all we need know about $\tilde{\mu}(z)$ is that, as a
consequence of the second law of thermodynamics, it must be what is termed a
positive real function \cite{ford88b}. That is, $\tilde{\mu}(z)$ must be
analytic with positive real part everywhere in the upper half plane and, in
addition, its boundary value on the real axis, which in general may be a
distribution, must satisfy the reality condition, 
\begin{equation}
\mathrm{Re}\{\tilde{\mu}(-\omega +i0^{+})\}=\mathrm{Re}\{\tilde{\mu}(\omega
+i0^{+})\}\geq 0.  \label{1.5}
\end{equation}
Using the expression (\ref{1.3}) for the correlation function, the mean
square displacement can be written 
\begin{equation}
s(t)=\frac{2\hbar }{\pi }\int_{0}^{\infty }d\omega \mathrm{Im}\{\alpha
(\omega +i0^{+})\}\coth \frac{\hbar \omega }{2kT}(1-\cos \omega t).
\label{1.6}
\end{equation}
It is convenient to consider instead of $s$ its time derivative 
\begin{equation}
\dot{s}(t)=\frac{2\hbar }{\pi }\int_{0}^{\infty }d\omega \mathrm{Im}\{\alpha
(\omega +i0^{+})\}\omega \coth \frac{\hbar \omega }{2kT}\sin \omega t.
\label{1.7}
\end{equation}
It should be clear from this formula that the \emph{long time} behavior
follows from the \emph{low frequency} behavior of the integrand. We now show
how this works for the most general case.

\section{Ohmic case.}

Consider first the Ohmic case, where $\tilde{\mu}(z)=\zeta $, a constant.
Then 
\begin{equation}
\alpha (\omega +i0^{+})=\frac{1}{-m\omega ^{2}-i\zeta \omega },  \label{2.1}
\end{equation}
and (\ref{1.7}) becomes 
\begin{equation}
\dot{s}(t)=\frac{2\hbar }{\pi }\int_{0}^{\infty }d\omega \frac{\zeta }{
m^{2}\omega ^{2}+\zeta ^{2}}\coth \frac{\hbar \omega }{2kT}\sin \omega t.
\label{2.2}
\end{equation}
The long time behavior is obtained by expanding the part of the integrand
other than the $\sin \omega t$ function in powers of $\omega $. That is, 
\begin{equation}
\dot{s}(t)\sim \frac{4kT}{\pi \zeta }\int_{0}^{\infty }d\omega \frac{\sin
\omega t}{\omega }=\frac{2kT}{\zeta }.  \label{2.3}
\end{equation}
Thus $\dot{s}$ is a constant and the mean square displacement is linear in
time. This is the case of so-called normal diffusion.

The case of zero temperature requires that in (\ref{2.2}) we set $\coth 
\frac{\hbar \omega }{2kT}\rightarrow 1$ first and then expand in powers of $\omega$,
to obtain 
\begin{equation}
\dot{s}(t)\sim \frac{2\hbar }{\pi \zeta }\int_{0}^{\infty }d\omega \sin
\omega t=\frac{2\hbar }{\pi \zeta t}.  \label{2.4}
\end{equation}
Note that this is a quantum result, vanishing in the limit $\hbar
\rightarrow 0$, while the finite temperature result (\ref{2.3}) is
classical. This is a general feature: diffusion (normal or anomalous) is
classical at finite temperature, quantum at zero temperature. The fact that
zero-temperature diffusion involves one power of $t$ less than for finite
temperature is also general.

It will be of some interest to exhibit the exact zero-temperature result,
which takes the form 
\begin{equation}
\dot{s}(t)=\frac{2\hbar }{\pi \zeta t}V(\frac{\zeta t}{m}),  \label{2.5}
\end{equation}
where \cite{bateman} 
\begin{equation}
V(x)=\int_{0}^{\infty }du\frac{x^{2}\sin (u)}{x^{2}+u^{2}}=\frac{x}{2}[e^{-x}
\mathrm{\bar{E}i}(x)-e^{x}\mathrm{Ei}(-x)].  \label{2.6}
\end{equation}
From its form, it should be clear that the
asymptotic form of the mean square displacement at zero temperature is given
by 
\begin{equation}
s(t)\sim \frac{2\hbar }{\pi \zeta }(\log \frac{\zeta t}{m}+\gamma_{E}),
\label{2.7}
\end{equation}
where $\gamma_{E}=0.577\cdots$ is Euler's constant.

\section{Colored noise.}

Consider the case where in the neighborhood of the origin 
\begin{equation}
\tilde{\mu}(z)=mb^{1-\gamma }(-iz)^{\gamma },\qquad -1<\gamma <1,
\label{3.1}
\end{equation}
where $b$ is a constant with the dimensions of frequency. It is easy to
verify that this is a positive real function if and only if $\gamma $ is
within the indicated range (we choose the branch with $-\pi <\arg (z)<\pi$
). With this, we see that (\ref{1.7}) becomes 
\begin{equation}
\dot{s}(t)=\frac{2\hbar }{\pi }\int_{0}^{\infty }d\omega \frac{\cos (\frac{
\pi }{2}\gamma )\coth \frac{\hbar \omega }{2kT}}{m\omega \lbrack (\frac{
\omega }{b})^{1-\gamma }+(\frac{b}{\omega })^{1-\gamma }+2\sin (\frac{\pi}{2
}\gamma )]}\sin \omega t.  \label{3.2}
\end{equation}
Expanding the integrand for small $\omega$, we find for large $t$, 
\begin{equation}
\dot{s}(t)\sim \frac{4kT\cos (\frac{\pi }{2}\gamma )}{\pi mb^{1-\gamma}}
\int_{0}^{\infty }d\omega \frac{\sin \omega t}{\omega ^{\gamma +1}}.
\label{3.3}
\end{equation}
Since 
\begin{equation}
\int_{0}^{\infty }dx\frac{\sin x}{x^{\gamma +1}}=\frac{\pi }{2\Gamma
(1+\gamma )\cos (\frac{\pi }{2}\gamma )},  \label{3.4}
\end{equation}
we have the result 
\begin{equation}
\dot{s}(t)\sim \frac{2kT}{mb^{1-\gamma }\Gamma (1+\gamma )}t^{\gamma }.
\label{3.5}
\end{equation}
Thus, we see that colored noise corresponds to anomalous diffusion. In
addition we see that there is a physical reason for the restricted range of $\gamma$:
it is a consequence of the second law of thermodynamics.

The case of zero temperature follows in the same way as in the Ohmic case.
Setting $\coth \frac{\hbar \omega }{2kT}\rightarrow 1$ and expanding in
powers of $\omega$, (\ref{3.2}) becomes 
\begin{equation}
\dot{s}(t)\sim \frac{2\hbar \cos (\frac{\pi }{2}\gamma )}{\pi mb^{1-\gamma }}
\int_{0}^{\infty }d\omega \frac{\sin \omega t}{\omega ^{\gamma }}=\frac{
\hbar \cot (\gamma \frac{\pi }{2})}{\Gamma (\gamma )mb^{1-\gamma }}t^{\gamma
-1}.  \label{3.6}
\end{equation}
Note that the case $\gamma =0$ is not singular ($\lim_{\gamma \rightarrow
0}\cot (\frac{\pi }{2}\gamma )/\Gamma (\gamma )=2/\pi$). Note also that
this zero-temperature result involves one power of $t$ less than the finite
temperature result (\ref{3.5}).

The extremes of the range of $\gamma $, $\gamma =\pm 1$, require special
attention. The case $\gamma =-1$ corresponds to a generalized susceptibility
of the form 
\begin{equation}
\alpha (z)=\frac{1}{-mz^{2}+mb^{2}}.  \label{3.7}
\end{equation}
This form corresponds to a free oscillator with force constant $K=mb^{2}$.
That is, $\gamma =-1$ corresponds to a harmonically bound particle. For this
case one can obtain from (\ref{1.7}) the exact result 
\begin{equation}
\dot{s}(t)=\frac{\hbar }{m}\coth \frac{\hbar b}{2kT}\sin bt.  \label{3.8}
\end{equation}
Thus, with $\gamma =-1$ the particle is bound and there is no diffusion.
This is consistent with (\ref{3.5}), for which the multiplier vanishes for $\gamma
=-1$.

The case $\gamma =+1$ in (\ref{3.1}) clearly corresponds to a shift in the
particle mass, with no dissipation. In the absence of dissipation, it should
be obvious that at finite temperature the mean square displacement should
grow with the square of the time, consistent with (\ref{3.5}) for this case.
There is, however, a non-trivial example for which dissipation is present
and the mean square displacement at finite temperature nevertheless grows as
the square of the time. This is the quantum electrodynamic (QED) case, which
we discuss next.

\section{QED case.}

In the case of a nonrelativistic electron coupled to the radiation field,
the Fourier transform of the memory function in QED can be written in the
form: \cite{ford85,ford98} 
\begin{equation}
\tilde{\mu}(z)=\frac{2e^{2}}{3c^{3}}\frac{z\Omega ^{2}}{z+i\Omega },
\label{4.1}
\end{equation}
where $\Omega$ is a large cut-off frequency, related to the electron
form-factor. With this, the form (\ref{1.4}) for the generalized
susceptibility for a free particle becomes 
\begin{equation}
\alpha (z)=\frac{z+i\Omega }{-mz^{3}-iM\Omega z^{2}},  \label{4.2}
\end{equation}
where $m$ is the bare mass and 
\begin{equation}
M=m+\frac{2e^{2}\Omega }{3c^{3}}  \label{4.3}
\end{equation}
is the renormalized (physical) mass. We can use $m$ as a parameter in place
of $\Omega$, writing 
\begin{equation}
\Omega =\frac{M-m}{M\tau _{e}},\qquad 0\leq m\leq M,  \label{4.4}
\end{equation}
where 
\begin{equation}
\tau _{e}=\frac{2e^{2}}{3Mc^{3}}\cong 6.25\times 10^{-24}\mathrm{s}.
\label{4.5}
\end{equation}
With this, we can write the generalized susceptibility in the form 
\begin{equation}
\alpha (z)=\frac{M-m-iMz\tau _{e}}{-Mz^{2}(M-m-imz\tau _{e})}.  \label{4.6}
\end{equation}
Note that with this, $m$ can be viewed as a measure of the strength of
coupling, with $0\leq m\leq M$. The limit $m=M$ corresponds to no
interaction (i.e., free particle) and the limit $m=0$ corresponds maximal
coupling, where the cut-off has its largest value consistent with causality (
$\Omega =\tau _{e}^{-1}$).

We can rearrange the form (\ref{4.6}) to write 
\begin{equation}
\alpha (z)=-\frac{1}{Mz^{2}}+\frac{(M-m)\tau _{e}}{-iMz(M-m-imz\tau _{e})}.
\label{4.7}
\end{equation}
Then we see that 
\begin{equation}
\mathrm{Im}\{\alpha (\omega +i0^{+})\}=-\frac{\pi }{M}\delta ^{\prime
}(\omega )+\frac{(M-m)^{2}\tau _{e}}{M\omega \lbrack (M-m)^{2}+(m\omega \tau
_{e})^{2}]},  \label{4.8}
\end{equation}
where $\delta ^{\prime }$ is the derivative of the delta-function. With
this, the expression (\ref{1.7}) for $\dot{s}$ becomes 
\begin{equation}
\dot{s}(t)=\frac{2kT}{M}t+\frac{2\hbar \tau _{e}}{\pi M}\int_{0}^{\infty
}d\omega \frac{(M-m)^{2}}{(M-m)^{2}+(m\omega \tau _{e})^{2}}\coth \frac{%
\hbar \omega }{2kT}\sin \omega t.  \label{4.9}
\end{equation}

\subsection{Free particle.}

Consider first the case of a free particle, with no interaction with the
radiation field. In this case only the first term in the expression (\ref
{4.9}) survives, and we have the result, 
\begin{equation}
s(t)=\frac{kT}{M}t^{2},\qquad \mathrm{free\ particle,}  \label{4.10}
\end{equation}
valid for all times. This result follows from equipartition ($\left\langle 
\frac{1}{2}Mv^{2}\right\rangle =\frac{1}{2}kT$) which for a free particle is
equally true in classical and quantum mechanics. Note that this is an exact
result: for a free particle at zero temperature $s(t)=0$ for all time.

\subsection{Maximal coupling.}

Next consider the case of maximal coupling, where $m=0$. In this case we
find the exact closed form result\cite{bateman}, 
\begin{eqnarray}
\dot{s}(t) &=&\frac{2kT}{M}t+\frac{2\hbar \tau _{e}}{\pi M}\int_{0}^{\infty
}d\omega \coth \frac{\hbar \omega }{2kT}\sin \omega t  \nonumber \\
&=&\frac{2kT}{M}(t+\tau _{e}\coth \frac{\pi kTt}{\hbar }).  \label{4.11}
\end{eqnarray}
Thus, for long time $\dot{s}$ grows linearly with time. Indeed the mean
square displacement for long times is exactly of the form (\ref{4.10}) for a
free particle. This is what we should expect since in QED a particle moving
with constant velocity feels no force.

At zero temperature, the exact expression (\ref{4.11}) becomes 
\begin{equation}
\dot{s}(t)=\frac{2\hbar \tau _{e}}{\pi M}\frac{1}{t}.  \label{4.12}
\end{equation}
Note that for long times this result is an exception to the general rule
that for colored noise the long time dependence at zero temperature involves
one power of $t$ less than at finite temperature. Here the difference is two
powers of $t$. This phenomenon is already signalled in the general
expression (\ref{3.6}) for colored noise at zero temperature, where the
multiplier vanishes for $\gamma =1$.

We note that at short times the exact expression (\ref{4.11}) diverges like $t^{-1}$.
This should be puzzling, since by it's very definition (\ref{1.1}) $s(0)=0$. This
divergence is a quantum phenomenon, arising from the mass renormalization. In order
to understand this, we consider next the general expression (\ref{4.9}) at zero
temperature.

\subsection{General case at zero temperature.}

In the limit of zero temperature, the general expression (\ref{4.9}) becomes 
\begin{equation}
\dot{s}(t)=\frac{2\hbar \tau _{e}}{\pi M}\int_{0}^{\infty }d\omega \frac{
(M-m)^{2}}{(M-m)^{2}+(m\omega \tau _{e})^{2}}\sin \omega t.  \label{4.13}
\end{equation}
For long times this vanishes like $t^{-1}$ (at zero temperature the particle
is at rest). For very short times, of order $m\tau _{e}/M$, this increases
rapidly from zero to a large value, after which it falls to zero again like $t^{-1}$.
In fact, this integral is exactly that appearing in the corresponding Ohmic case
(\ref{2.5}). We can therefore write 
\begin{equation}
\dot{s}(t)=\frac{2\hbar \tau _{e}}{\pi Mt}V(\frac{(M-m)t}{m\tau _{e}}),
\label{4.14}
\end{equation}
where $V(x)$ is given by (\ref{2.6}). This is an example in which the bare
mass appears for extremely short times (i.e., extremely high frequencies)
Another example is the canonical commutation rule \cite{ford89}. Clearly, in
the limit $m\rightarrow 0$, (\ref{4.14}) becomes exactly of the form (\ref
{4.12}), but now with $\frac{1}{t}$ interpreted as the principal value. As
in the Ohmic case, in order to integrate to find the asymptotic form of the
mean square displacement we need to know the behavior at small $t$. With the
result (\ref{4.14}) we see that 
\begin{equation}
s(t)\sim \frac{2\hbar \tau _{e}}{\pi M}\log \frac{(M-m)t}{m\tau _{e}}+\gamma_{E}.
\label{4.15}
\end{equation}
Thus, we see that the bare mass appears even in the long time behavior of $s(t)$. We
shall have more to say about this in our discussion of wave packet spreading.

\section{Spreading of a wave packet.}

As a somewhat different application of these results, we consider the
spreading of a Gaussian wave packet. The situation is as follows. At time $t_{1}$ the
Brownian particle, which before the measurement is in equilibrium at temperature $T$,
is measured (i.e., detected) with a Gaussian instrument of width $\sigma _{1}$ and
centered at $x_{1}$. Then at a later time $t_{2}$ a second measurement is made, again
with a Gaussian instrument but now with width $\sigma _{2}$ and centered at $x_{2}$.
Let $W(1,2)dx_{1}dx_{2}$ be the probability that the first measurement finds the
particle in range $dx_{1}$ about $x_{1}$ \emph{and} that the second measurement finds
the particle in range $dx_{2}$ about $x_{2}$. The mean square width of the wave
packet at time $t_{2}$ is then given by 
\begin{equation}
w^{2}(t_{2}-t_{2})\equiv \int_{-\infty }^{\infty }dx_{1}\int_{-\infty
}^{\infty }dx_{2}W(1,2)(x_{1}-x_{2})^{2}.  \label{5.1}
\end{equation}
Here we must be sure to distinguish this expectation from that in (\ref{1.1}
); here $x_{1}$ and $x_{2}$ are instrumental parameters (i.e., c-numbers) 
\emph{not} quantum mechanical operators. One could well argue that for a
quantum particle this wave packet width $w^{2}$ is a better measure of
anomalous diffusion than the mean square displacement $s$. The picture, to
repeat, is the following. An initial wave packet of width $\sigma _{1}$ is
formed at time $t_{1}$. In the course of time this wave packet will spread.
It's center will not move since the mean velocity is zero. At a later time $t_{2}$
the mean square width, \ including the width $\sigma _{2}$ of the second instrument,
is given by (\ref{5.1}).

Now, for quantum Brownian motion an exact expression for $W(1,2)$ can be
obtained \cite{ford86}, 
\begin{equation}
W(1,2)=\frac{1}{2\pi \sigma \tau \sqrt{1-\rho ^{2}}}\exp \{-\frac{1}{
2(1-\rho ^{2})}(\frac{x_{1}^{2}}{\sigma ^{2}}-2\frac{\rho x_{1}x_{2}}{\sigma
\tau }+\frac{x_{2}^{2}}{\tau ^{2}})\},  \label{5.2}
\end{equation}
where (note a misprint in Eq. 7.18 of \cite{ford86}) 
\begin{eqnarray}
\sigma ^{2} &=&\sigma _{1}^{2}+\left\langle x(t_{1})^{2}\right\rangle 
\nonumber \\
\tau ^{2} &=&\sigma _{2}^{2}+\left\langle x(t_{1})^{2}\right\rangle -\frac{
[x(t_{1}),x(t_{2})]^{2}}{4\sigma _{1}^{2}},  \nonumber \\
\sigma \tau \rho &=&\frac{1}{2}\left\langle
x(t_{1})x(t_{2})+x(t_{2})x(t_{1})\right\rangle .  \label{5.3}
\end{eqnarray}
With this, we find that (\ref{5.1}) becomes \cite{ford01}
\begin{equation}
w^{2}(t_{2}-t_{1})=s(t_{2}-t_{1})+\sigma _{1}^{2}-\frac{
[x(t_{1}),x(t_{2})]^{2}}{4\sigma _{1}^{2}}+\sigma _{2}^{2}.  \label{5.4}
\end{equation}
Thus, the mean square width of the wave packet is a sum of four terms. The
first is the mean square displacement (\ref{1.1}). The second and third
terms involve the finite width of the initial wave packet, with the third
term clearly arising from the uncertainty principle. Finally, the last term
is the mean square width of the second instrument. Clearly, this final width
can be taken to be zero, but not the initial width.

For quantum Brownian motion, the commutator is a c-number given by \cite{ford88b} 
\begin{equation}
\lbrack x(t_{1}),x(t_{2})]=\frac{2i\hbar }{\pi }\int_{0}^{\infty }d\omega 
\mathrm{Im}\{\alpha (\omega +i0^{+})\}\sin \omega (t_{2}-t_{1}).  \label{5.5}
\end{equation}
Note that the commutator is temperature independent. We now consider a
number of cases.

\subsection{Colored noise}

For long times, using the method described above, we find for colored noise, 
\begin{equation}
\lbrack x(t_{1}),x(t_{2})]\sim \frac{i\hbar }{mb^{1-\gamma }\Gamma (1+\gamma
)}t^{\gamma },  \label{5.6}
\end{equation}
where to save writing we have written $t=t_{2}-t_{1}$. At finite
temperature, using this result for the commutator and the corresponding
result (\ref{3.5}) for the mean square displacement, we see that for $-1<\gamma <1$
and for sufficiently long times the term $s(t_{2}-t_{1})$ dominates in (\ref{5.4}).
Thus at finite temperature, the long time behavior for colored noise is the same,
whether measured by the mean square displacement (\ref{1.1}) or by the wave packet
width (\ref{5.1}), and always classical.

On the other hand, at zero temperature, using the result (\ref{3.6}) we see
that the general expression (\ref{5.4}) becomes 
\begin{eqnarray}
w^{2}(t) &\sim &\frac{\hbar \cot (\gamma \frac{\pi }{2})}{\Gamma (1+\gamma
)mb^{1-\gamma }}t^{\gamma }  \nonumber \\
&&+\sigma _{1}^{2}+\frac{\hbar ^{2}}{4m^{2}\sigma _{1}^{2}b^{2-2\gamma
}\Gamma (1+\gamma )^{2}}t^{2\gamma }+\sigma _{2}^{2}.  \label{5.7}
\end{eqnarray}
Here the first term corresponds to $s(t)$ and we see that for $-1<\gamma <0$
this term still dominates at zero temperature. The Ohmic case $\gamma =0$
appears to be singular, but as we noted above this is not a problem. In
fact, for the Ohmic case at zero temperature we have 
\begin{equation}
w_{\mathrm{Ohmic}}^{2}(t)\sim \frac{2\hbar }{\pi \zeta }\log \frac{\zeta t}{m
}+\sigma _{1}^{2}+\frac{\hbar ^{2}}{4\sigma _{1}^{2}\zeta ^{2}}+\sigma
_{2}^{2}.  \label{5.8}
\end{equation}
Thus, in the Ohmic case, we see that the first term still dominates. On the
other hand, for $0<\gamma <1$, for sufficiently long times the third term
dominates, corresponding to what we might call uncertainty principle
spreading.

\subsection{Free particle}

For the free particle, using the expression (\ref{4.8}) with $m=M$, the
expression (\ref{5.5}) for the commutator becomes 
\begin{equation}
\lbrack x(t_{1}),x(t_{2})]=\frac{i\hbar }{M}(t_{2}-t_{1}).  \label{5.9}
\end{equation}
With this and the expression (\ref{4.10}) for the mean square displacement,
we see that the general expression (\ref{5.4}) becomes 
\begin{equation}
w_{\mathrm{free}}^{2}(t)=\frac{kT}{M}t^{2}+\sigma _{1}^{2}+\frac{\hbar
^{2}t^{2}}{4M^{2}\sigma _{1}^{2}}+\sigma _{2}^{2}.  \label{5.10}
\end{equation}
First of all, we note that at zero temperature and with $\sigma _{2}^{2}=0$
this becomes exactly the formula for wave packet spreading found in
textbooks \cite{schiff}. For finite temperature the first term, the mean
square displacement, has the same $t^{2}$ dependence as the third term, the
uncertainty principle spreading. The ratio of these two terms is $\lambda
^{2}/8\pi \sigma _{1}^{2}$, where $\lambda =\sqrt{2\pi \hbar ^{2}/MkT}$ is
the mean thermal de Broglie wavelength. Thus, we see that the spreading will
be classical provided that $\lambda \ll \sigma _{1}$.

\subsection{QED}

As in our discussion of colored noise, the case $\gamma =1$ requires special
consideration and we discuss it in terms of the QED case. For the general
form (\ref{4.8}) of $\mathrm{Im}\{\alpha \}$, we see that (\ref{5.5})
becomes 
\begin{equation}
\lbrack x(t_{1}),x(t_{2})]_{\mathrm{QED}}=\frac{i\hbar }{M}\{t+\frac{2}{\pi }
\int_{0}^{\infty }d\omega \frac{(M-m)^{2}\tau _{e}\sin \omega t}{\omega
\lbrack (M-m)^{2}+(m\omega \tau _{e})^{2}]}\}.  \label{5.11}
\end{equation}
The integral appearing here is standard \cite{bateman}, and we find 
\begin{equation}
\lbrack x(t_{1}),x(t_{2})]_{\mathrm{QED}}=\frac{i\hbar }{M}\{t+\tau
_{e}(1-e^{-(M-m)t/m\tau _{e}})\}.  \label{5.12}
\end{equation}
Here the exponential is exceedingly small for any time of the order of $\tau
_{e}$ or longer. It does, however, guarantee that the canonical equal-time
commutator, $[x(t),\dot{x}(t)]_{\mathrm{QED}}=i\hbar /m$, does involve the
bare mass, as it must \cite{ford89}.

With this result, it should be clear that for long times the leading
behavior of the wave packet width in QED is the same as that for a free
particle. There will be lower order corrections, of course, but these are
perhaps best discussed in terms of the zero temperature result. At zero
temperature, using the results (\ref{4.15}) and (\ref{5.12}) we find 
\begin{equation}
w_{\mathrm{QED}}^{2}(t)\sim \frac{2\hbar \tau _{e}}{\pi M}\{\log \frac{(M-m)t
}{m\tau _{e}}+\gamma_{E} \}+\sigma _{1}^{2}+\frac{\hbar ^{2}(t+\tau _{e})^{2}}{
4M^{2}\sigma _{1}^{2}}+\sigma _{2}^{2}.  \label{5.13}
\end{equation}

\section{Conclusions}

The first calculation of normal diffusion [equivalent to our equation (2.3)] was carried
out by Einstein in his famous explanation \cite{einstein} for the observations of Brown
\cite{brown} on the  random motion of pollen grains immersed in a fluid.  The term
"Brownian motion" is now used in a generic sense to denote random motion and it covers
a wide spectrum of phenomena.  Einstein's explanation of the original
Brownian motion used a diffusion equation but, shortly afterwards,
Langevin \cite{langevin} presented an entirely new approach to the
subject, based on the introduction of a stochastic differential equation,
which in the words of Chandrasekhar \cite{chandrasekhar}, constitutes the
"modern" approach to this and other such problems.  Langevin's analysis was
confined to the classical long-time motion of a free particle at high
temperature in an Ohmic heat bath (based on Stokes analysis of viscous drag).

Here, we use a generalized Langevin equation which incorporates quantum and
non-Markovian (memory) effects, arbitrary times, arbitrary temperature and a
very general heat bath.  This leads to a fluctuation-dissipation theorem
which enables us to calculate anomalous (in a sense that it goes beyond the
Einstein-Langevin case) diffusion for a wide variety of parameters which, in
turn, will facilitate analysis of a variety of physical phenomena.  In
particular, we applied our results to wave-packet spreading.  We consider the use of
the quantum Langevin equation and the fluctuation-dissipation theorem to be simpler and
more physically appealing than approaches based on path-integral techniques
\cite{grabert87,hakim85}.  In addition, our description of the motion of a quantum
particle in a linear passive heat bath is characterized by $Re[\tilde{\mu}(z)]$, the
Fourier transform (spectral distribution) of the memory function appearing in the
quantum Langevin equation.  As a consequence of the second law of thermodynamics,
$\tilde{\mu}(z)$ must be a positive real function \cite{ford88b}, that is
$\tilde{\mu}(z)$ must be analytic with positive real part everywhere in the upper half
plane and, in addition, its boundary value on the real axis, which in general may be a
distribution, must satisfy the reality condition, given by (1.5).  As a consequence,
this limits the forms of $\tilde{\mu}(z)$, as exemplified in (3.1), which in turn
limits the forms of anomalous diffusion.  Finally, we discussed in detail the special
case of an electron interacting with the radiation field.

\acknowledgments

We wish to thank the School of Theoretical Physics, Dublin Institute for
Advanced Studies, for their hospitality.


\begin{thebibliography}{99}
\bibitem{klafter} J. Klafter and I.M. Sokolov, Phys. World Aug. 2005, p.29.

\bibitem{andrade} M.F. de Andrade et al., Phys. Lett. A \textbf{347}, 160 (2005).

\bibitem{lutz}  E. Lutz, Phys. Rev. E \textbf{64}, 051106 (2001).

\bibitem{tsallis} C. Tsallis and D.J. Bukman, Phys. Rev. E  \textbf{54},
R2197 (1996).

\bibitem{scullya} M. O. Scully, G. Sussmann and C. Benkert, Phys. Rev.
Lett. \textbf{60}, 1014 (1988)

\bibitem{scully} M. O. Scully, M. S. Zubairy and K. Wodkiewicz, Optics Comm. 
\textbf{65}, 440 (1988)

\bibitem{benkert} C. Benkert, M.O. Scully, A.A. Rangwala and W. Schleich,
Phys. Rev. A \textbf{43}, 1503 (1990).

\bibitem{schaufler}  S. Schaufler, W.P. Schleich and V.P. Yakovlev, Phys.
Rev. Lett. \textbf{83}, 3162 (1999).

\bibitem{grabert87} H. Grabert, P. Schramm and G.-L. Ingold, Phys. Rev.
Lett. \textbf{58}, 1285 (1987).

\bibitem{callen} H.B. Callen and T.A. Welton, Phys. Rev. \textbf{83}, 34 (1951).

\bibitem{ford88a}  G.W. Ford, J.T. Lewis and R.F. O'Connell, Ann. Phys. (NY) 
\textbf{185}, 270 (1988).

\bibitem{ford88b} G.W. Ford, J.T. Lewis and R.F. O'Connell, Phys. Rev. A  
\textbf{37}, 4419 (1988).

\bibitem{bateman}  \textit{Tables of integral transforms}, edited by Staff of the 
Bateman Manuscript Project, Bateman manuscript project, Vol. 1 (McGraw-Hill, 
New York, 1954).

\bibitem{ford85}  G.W. Ford, J.T. Lewis and R.F. O'Connell, Phys. Rev. Lett. 
\textbf{55}, 2273 (1985).

\bibitem{ford98} G.W. Ford and R.F. O'Connell, Phys. Rev. A \textbf{57},
3112 (1998).

\bibitem{ford89} G.W. Ford and R.F. O'Connell, Journ. Stat. Phys. \textbf{57}
(1989) 803.

\bibitem{ford86} G. W. Ford and J. T. Lewis, \emph{Probability, Statistical
Mechanics, and Number Theory}, Advances in Mathematics Supplemental Studies,
Volume 9 (Academic Press, Orlando, Florida 1986)

\bibitem{ford01} G.W. Ford, J.T. Lewis and R.F. O'Connell, Phys. Rev. A \textbf{64},
032101 (2001).

\bibitem{schiff} L. I. Schiff, \emph{Quantum Mechanics} (McGraw-Hill, New
York 1949) p. 58.

\bibitem{einstein} A. Einstein, \emph{Eine neue Bestimmung der
Molek$\ddot{u}$ldimensionen}, (Wyss, Bern, 1905) [Einstein's thesis];
ibid. Ann. d. Physik \textbf{17}, 549 (1905); this and the further
articles of Einstein appear in \emph{Investigations on the Theory of the Brownian
Movement} (Dover, New York, 1956). 

\bibitem{brown} R. Brown, Philos. Mag. \textbf{4}, 161 (1828);
\emph{Ibid}. \textbf{6}, 161 (1829).

\bibitem{langevin} M. P. Langevin, C. R. Acad, Sci., Paris \textbf{146},
530 (1908)  (a translation of this article appears in D.S. Lemons, A.
Gythiel, Am. J. Phys. \textbf{65} 1079 (1997)).

\bibitem{chandrasekhar} S. Chandrasekhar, Rev. Mod. Phys. \textbf{15}, 1
(1943). 

\bibitem{hakim85} V. Hakim and V. Ambegaokar, Phys. Rev. A \textbf{32}, 423 (1985). 
So far as we know, this paper contains the first derivation of the width
of a spreading wave packet, equivalent with our Eq. (5.4), but only for
the ohmic case. Also, we note a misprint in their expression (38) for
the width. 

\end{thebibliography}
\end{document}